\documentclass[diagnostics,article,submit,pdftex,moreauthors]{Definitions/mdpi} 
\firstpage{1} 
\pubvolume{1}
\issuenum{1}
\articlenumber{0}
\pubyear{2026}
\copyrightyear{2026}
\datereceived{ } 
\daterevised{ } 
\dateaccepted{ } 
\datepublished{ } 

\usepackage{xcolor}

\Title{Risk Prediction in Cancer Imaging Using Enriched Radiomics Features}

\Author{Alec Reinhardt$^{1,\dagger}$, Tsung-Hung Yao$^{1,\dagger}$\orcidA{}, Raven Hollis$^{2}$, Galia Jacobson$^{2}$, Millicent Roach$^{2}$, Mohamed Badawy$^{2}$, Peter Park$^{2}$, Laura Beretta$^{3}$,David Fuentes$^{4}$,Newsha Nikzad$^{5}$,Prasun Jalal$^{5}$,Eugene Koay$^{2,*}$ and Suprateek Kundu$^{1,*}$}

\AuthorNames{Alec Reinhardt, Tsung-Hung Yao, Raven Hollis, Galia Jacobson, Millicent Roach, Mohamed Badawy, Peter Park, Laura Beretta, David Fuentes, Newsha Nikzad, Prasun Jalal, Eugene Koay and Suprateek Kundu}

\address{%
$\dagger$ \quad Alec Reinhardt and Tsung-Hung Yao contributed equally to this work\\
$^{1}$ \quad Department of Biostatistics, The University of Texas MD Anderson Cancer Center, TX, USA\\
$^{2}$ \quad Department of Radiation Oncology, The University of Texas MD Anderson Cancer Center, TX, USA\\
$^{3}$ \quad Department of Molecular and Cellular Oncology, The University of Texas MD Anderson Cancer Center, TX, USA\\
$^{4}$ \quad Department of Imaging Physics, The University of Texas MD Anderson Cancer Center, TX, USA\\
$^{5}$ \quad Division of Gastroenterology and Hepatology, The Baylor College of Medicine Medical Center, TX, USA\\}

\corres{Correspondence: EKoay@mdanderson.org (E.K.) and skundu2@mdanderson.org (S.K.)}

\abstract{{\noindent \textbf{Background/Objectives}:} We aim to develop enriched radiomics features that integrate classical structural radiomics with novel functional radiomics derived from liver MRI for diagnosis and risk stratification in liver cancer. The proposed framework leverages enhancement pattern mapping (EPM) images to provide an automated and robust radiomics representation that captures intratumoral heterogeneity through pixel-level functional information. \textbf{Methods}: Pixel-wise EPM data reflecting blood perfusion were extracted from T1-weighted MRI scans. Classical structural radiomics features were extracted via existing software such as PyRadiomics. In addition, empirical quantiles of EPM values over all pixels within the image, and then smoothed using suitable basis. The smoothed quantiles, along with the classical structural quantiles, are used as functional radiomics features for diagnostic classification and tumor grade stratification, using L1-penalized logistic regression that automatically downweights the contribution of the irrelevant features. To further characterize disease progression, we conducted longitudinal analyses using Bayesian tensor response regression (BTRR), which enables spatial smoothing and parsimonious modeling of temporally evolving imaging patterns. \textbf{Results:} The enriched radiomics features illustrate higher diagnostic classification performance (AUC=0.96, sensitivity$\ge 0.8$) and superior tumor grade stratification accuracy (AUC=0.87, sensitivity=0.8) compared to alternate radiomics features. Moreover, we find that the proportion of lesion pixels with significant reduction in EPM values over time is considerably higher (median $\approx$ 0.12) in aggressive lesions versus stable or mildly aggressive lesions (median $\approx$ 0.025). \textbf{Conclusion:} The enriched novel radiomics features have potential to replace classical radiomics analysis and to be used as imaging biomarkers in cross-sectional cancer imaging analysis and longitudinal cancer imaging studies.}


\begin{document}




\section{Introduction}

\label{sec:introduction}
Hepatocellular carcinoma (HCC) is the most common type of primary liver cancer and a significant global health concern. It is the sixth most common cancer and the fourth leading cause of cancer-related deaths worldwide. One of the key factors contributing to its high mortality rate is the lack of 
diagnosis of advanced stage for the majority of patients. Early detection of HCC may plays a pivotal role in improving liver cancer prognosis and increasing treatment success rates. 
Patients diagnosed at an early stage have the opportunity to undergo curative treatments such as resection, ablative therapies, and liver transplantation, while those diagnosed at a later stage typically only qualify for palliative systemic treatments with limited effectiveness. Consequently, the 5-year survival rate surpasses 70\% for individuals with early-stage HCC, whereas it falls below 5\% for those diagnosed at advanced stages. Surveillance for HCC is recommended in at-risk patients, including those with cirrhosis of all etiologies, and certain populations with chronic Hepatitis B virus infection. Guidelines recommend HCC surveillance with abdominal ultrasound (US) with or without serum alpha-fetoprotein measurement \cite{yildirim2023advances} every 6 months. Other diagnostic imaging modalities for surveillance include computed tomography (CT) and magnetic resonance imaging (MRI), and ultrasound imaging. However, the sensitivity of the surveillance may potentially vary with the imaging modality \cite{tzartzeva2018surveillance}.

Although medical imaging has made rapid advances and has emerged as one of the most promising tools for early diagnosis and risk detection, there are important limitations. These include both MRI-based over surveillance due to the frequent detection of benign vascular lesions, and delayed HCC diagnosis due to the under-staging of small or early malignant lesions in approximately 25\% of cases \cite{mahmud2020risk, park2020enhancement}. A potential reason for these inadequacies is the inherent noise in liver MRI scans, which is often unaccounted for. Another major diagnostic and prognostic challenge of rising concern is the heterogeneity of pre-cancerous and cancerous lesions. Despite technological advances, lesion detection, characterization of patterns of enhancement, and monitoring remain sub-optimal. While the Liver Imaging Reporting and Data System (LIRADS) has been developed as a standardized classification system for liver nodules (depending on lesion size, arterial phase enhancement, capsular enhancement, washout, and threshold growth to deduce a nodule's probability of malignancy, and so on), risk stratification is hindered by significant heterogeneity within LIRADS categories, especially for LR-3 and LR-4 lesions of which 38\% and 74\% are HCC, respectively \cite{chernyak2018liver, nikzad2022characterization}. A recent meta-analysis showed that there was significant heterogeneity in sensitivity of US-based surveillance (21\%-89\%) across studies included in the meta-analysis \cite{tzartzeva2018surveillance}, which can be potentially attributed to the above factors and highlights a limitation of imaging-based surveillance. Furthermore, in other domains such as pancreatic cancer, standardized scoring system (such as LIRADS in Liver cancer) may not always be available. This exacerbates difficulties for imaging-based risk stratification, and the standard of care may resort to biopsy or other invasive procedures for a definitive diagnostic assessment. Therefore, there is a critical need for new minimally invasive imaging platforms and accompanying statistical approaches that can render risk stratification with high sensitivity and specificity based on imaging data collected at earlier disease stages.

Radiomics has emerged as a tool for quantifying solid tumor phenotype through the extraction of quantitative radiographic features \cite{Gillies2016-qm}. There is a growing body of evidence pointing to the prognostic value of such features \cite{Aerts2014-gs, Ganeshan2010-eg}. In the context of HCC, radiomics-based approaches have been applied to predict the emergence of new HCC lesions and to support risk stratification in high-risk cirrhotic populations. In particular, \cite{Tietz2021} developed a CT-based structural radiomics framework to predict the development of new HCC lesions using handcrafted intensity and texture features. While such studies demonstrate the promise of quantitative imaging biomarkers for HCC surveillance, they primarily rely on structural descriptors derived directly from imaging intensities and do not explicitly characterize the distributional properties of voxel-level enhancement patterns. While radiomics features are often hand-crafted \cite{Wu2016-is} some recent studies have adapted deep learning based feature extraction where representative features are learned automatically from data \cite{Hosny2019-jy, Cho2021-ih}. Unfortunately, neural network-based automated feature extraction approaches often lack interpretability. 

In order to address this gap in literature, we develop a distributional data analysis (DDA) approach that extracts enriched probabilistic radiomics features to develop predictive models in cancer imaging. In particular, we develop interpretable distributional-based features depending on quantile distributions that leverages the information embedded in pixel-level imaging measurements across the entire organ (or lesional area). Therefore, these probabilistic features provide detailed granular information for modeling clinical outcomes. By utilizing full information embedded in the probability distribution of granular pixel-level measurements, the proposed approach is able to account for the heterogeneity within the image that may arise from the differences of histopathological properties of lesional, peri-lesional and healthy tissues. Our use of quantile distributions for risk prediction leverages the recent literature on DDA literature that has gained increasingly popularity in literature for analysis involving covariates with their own probability distributions \cite{augustin2017modelling, dumuid2020compositional, ghodrati2022distribution}. These novel probabilistic features compliment the traditional structural radiomics features that are used in classical radiomics pipelines \cite{Aerts2014}. 

Subsequently, we combine the probabilistic radiomics features with the classical structural radiomics features under a penalized logistic regression model to provide an enriched predictive analysis of the liver cancer imaging dataset. In particular, we leverage enhancement pattern mapping (EPM) images \cite{park2020enhancement} derived from multi-phase T1-MRI scans of the liver, in order to perform diagnosis and LIRADS lesion score stratification. The EPM images were chosen based on evidence in the previous literature \cite{park2020enhancement} that illustrates that EPM derived from CT images improves the contrast-to-noise (CNR) ratio enhancement for lesion detection in hepatobiliary malignancy. We used the EPM images to derive classical (structural) and novel functional (quantile-based) radiomic characteristics, and subsequently used these characteristics to predict clinical diagnosis, as well as LIRADS score. We illustrate considerably improved prediction ability of the enriched (involving both classical/structural and quantile-based) radiomic features in prediction tasks compared to prediction based on only one type of radiomic feature. In addition to evaluating the predictive potential of the EPM features, we also investigate the potential of these images as neuroimaging biomarkers by evaluating significant longitudinal changes in the EPM values within the lesional and peri-lesional areas. Our analysis reveals that aggressive tumors that deteriorate longitudinally typically contain a larger proportion of imaging pixels in and around the lesion with significant reduction in EPM values over visits, compared to stable tumors. This finding is expected to have useful translational impact for longitudinal cancer imaging studies relying on EPM images. Collectively, these combined analyses establish a comprehensive prediction framework based on pixel-level EPM imaging data, capable of addressing gaps in literature.

\begin{figure}[h]
    \centering
    \includegraphics[width=\textwidth]{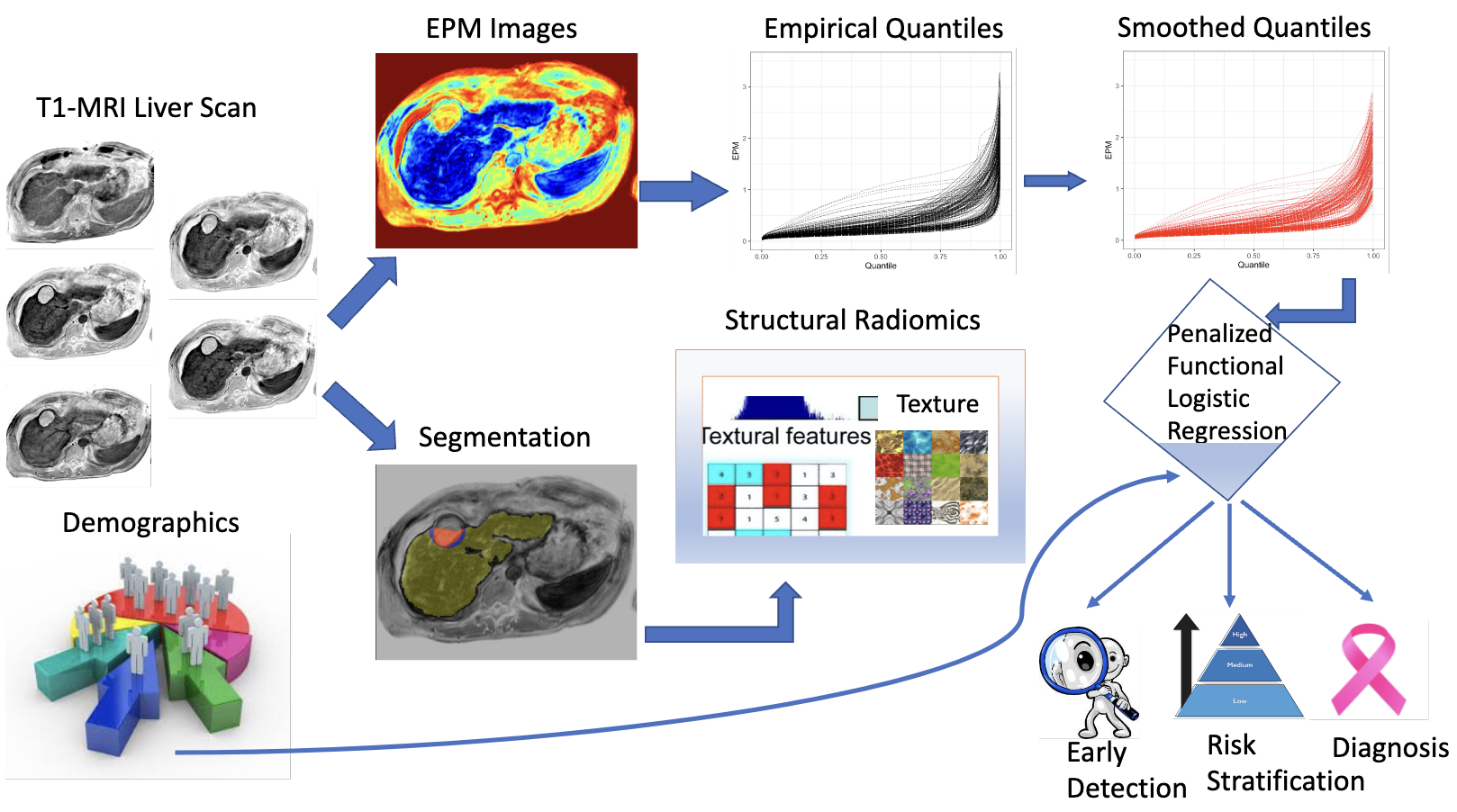}
    \caption{A schematic of the proposed analytical pipeline. The leftmost panel captures the 5 different phases of the T1-MRI liver scan that are all registered to a common template for a given subject. Subsequently the EPM images are computed from these T1-MRI scans, and then empirical quantile distributions followed by smoothed quantile distributions (based on quantlet basis expansions) were computed. The arterial phase of T1-MRI is used for segmentation of tumor and also used to compute the structural radiomics features. These two types of features, along with demographics are used for three classification tasks that include diagnosis and lesion score stratification.}
    \label{fig:schema}
\end{figure}

\section{Materials and Methods}

\subsection{Description of Liver Imaging Dataset}
The study cohort comprised patients with cirrhosis who presented to a tertiary care Hepatology Clinic between 2012 and 2020 and were prospectively followed under an IRB-approved protocol (PA14-0646) with a waiver of informed consent. These individuals underwent biannual contrast-enhanced MRI for hepatocellular carcinoma (HCC) surveillance. Inclusion criteria required at least two consecutive liver-protocol MRIs. Patients with HCC diagnosed at baseline were excluded.

The final dataset included 97 control subjects (without lesions) and 48 case subjects with LR-3 or LR-4 lesions. After excluding some case subjects and visits based on EPM image or segmentation quality, 38 cases remained, 35 of whom had imaging data at two timepoints. Table \ref{tab:Summary_Subj} provides descriptive statistics for both groups. Across the 73 case images (35 subjects $\times$ 2 visits $+$ 3 subjects $\times$ 1 visit), 88 distinct lesion regions were identified, excluding tiny lesions with fewer than 10 pixels. These included both pre- and post-diagnostic lesions. Among the 35 subjects with paired scans, changes in lesion grade and characteristics were observed over time. Each lesion had a corresponding LIRADS score (ranging from 2 to 5), assigned using 2018 LIRADS criteria, based on arterial phase hyperenhancement, venous/delayed washout, lesion size, and growth dynamics. Spatially distinct lesions were delineated via a clustering algorithm, separating them if the closest points were $\geq$5 mm apart. Table S1 in Supplementary Materials summarizes lesion characteristics.

\begin{table}[h]
    \centering
    \begin{tabular}{lccc}
    \toprule
        &\textbf{ Case} &\textbf{ Control} &\textbf{ Overall}\\
        &\textbf{ (N=35)} &\textbf{ (N=97)} &\textbf{ (N=132)}\\
    \midrule
   \textbf{ Mean EPM in Liver} & & &\\
    \quad Mean (SD) & 0.431 (0.160) & 0.354 (0.119) & 0.374 (0.135) \\
    \quad Median [Min, Max] & 0.406 [0.205, 0.953] & 0.379 [0.167, 0.660] & 0.386 [0.167, 0.953] \\
    \addlinespace
   \textbf{ Age} & & &\\
    \quad Mean (SD) & 57.9 (9.17) & 58.9 (7.78) & 58.7 (8.15) \\
    \quad Median [Min, Max] &  60.0 [23.0, 73.0] & 59.0 [26.0, 83.0] & 60.0 [23.0, 83.0] \\
    \quad Missing & 0 (0\%) & 1 (1.0\%) & 1 (0.8\%)\\
    \addlinespace
   \textbf{ Sex} & & &\\
    \quad F & 12 (34.3\%) & 36 (37.1\%) & 48 (36.4\%) \\
    \quad M & 23 (65.7\%) & 60 (61.9\%) & 83 (62.9\%) \\
    \quad Missing & 0 (0\%) & 1 (1.0\%) & 1 (0.8\%)\\
    \addlinespace
   \textbf{ Race} & & &\\
    \quad White & 28 (80.0\%) & 87 (89.7\%) & 115 (87.1\%) \\
    \quad Asian & 3 (8.6\%) & 1 (1.0\%) & 4 (3.0\%) \\
    \quad Black & 2 (5.7\%) & 6 (6.2\%) & 8 (6.1\%) \\
    \quad Other & 2 (5.7\%) & 2 (2.1\%) & 4 (3.0\%) \\
    \quad Missing & 0 (0\%) & 1 (1.0\%) & 1 (0.8\%) \\
    \addlinespace
   \textbf{ Ethnicity} & & &\\
    \quad Hispanic & 8 (22.9\%) & 26 (26.8\%) & 34 (25.8\%) \\
    \quad Non Hispanic & 27 (77.1\%) & 70 (72.2\%) & 97 (73.5\%) \\
    \quad Missing & 0 (0\%) & 1 (1.0\%) & 1 (0.8\%) \\
    \addlinespace
   \textbf{ BMI} & & &\\
    \quad Mean (SD) & 30.4 (5.25) & 30.9 (6.13) & 30.8 (5.89) \\
    \quad Median [Min, Max] &  30.9 [20.0, 41.0] & 31.2 [18.9, 48.4] & 31.1 [18.9,48.4] \\
    \quad Missing & 0 (0\%) & 1 (1.0\%) & 1 (0.8\%)\\
    \addlinespace
   \textbf{ Diabetes} & & &\\
    \quad No & 26 (74.3\%) & 65 (67.0\%) & 91 (68.9\%) \\
    \quad Yes & 9 (25.7\%) & 31 (32.0\%) & 40 (30.3\%) \\
    \quad Missing & 0 (0\%) & 1 (1.0\%) & 1 (0.8\%)\\
    \bottomrule 
    \end{tabular}
    \caption{Summary of demographic and liver image information of patients.}
    \label{tab:Summary_Subj}
\end{table}

\subsubsection{Enhancement Pattern Mapping (EPM) Images}
The 3-D voxel-based EPM approach was earlier used for quantitative image analysis of liver lesions based on CT scans \cite{park2020enhancement}, which we now adapt for MRI scans. The original T1w-MRI scans were processed to obtain voxel-wise EPM data that corresponds to the blood perfusion in the liver tissue. The generalized enhancement pattern, such as quantitative intensity changes throughout multi-phase MRI due to uptake and washout of contrast materials was acquired from the registered multi-phase MRI scans. A normal liver enhancement curve was developed through fitting the MRI intensity values over time within user-selected ROIs, sampled uniformly across cirrhotic liver parenchyma from the same patient. The root-mean-square deviation (RMSD) for each voxel was computed by averaging the squares of differences between the generalized normal liver intensity and the voxel intensity across all time points, and subsequently taking the square root of the average. The RMSD values of each voxel was mapped to their original MRI coordinates, and these values comprised the EPM image maps used for our analysis. The normal liver enhancement curve was obtained by fitting MRI intensity values sampled from the normal liver ROIs over the period of contrast phases by a piece-wise smooth function, where each piece was a second-order polynomial. The EPM algorithm was implemented numerically using MATLAB.

\subsection{Extraction of Radiomics Features} 
We engineer and extract novel functional radiomics features from the EPM images, and combine them with classical structural radiomics features from the MRI scans in our analysis. 

\begin{figure}
    \centering
    \includegraphics[width=.8\textwidth]{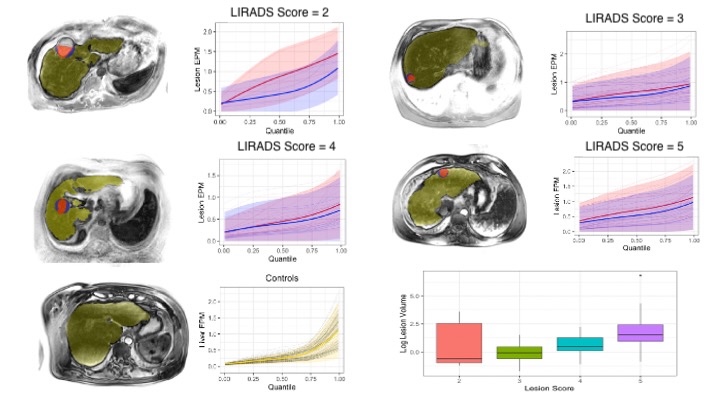}
    \caption{\label{fig:Fig2} Visualizations of lesion EPM images with segmentations for LIRADS scores 2 to 5 and healthy control. Liver masks are shown in yellow; lesion masks in red and surrounding peri-lesional areas are shaded in blue. Mean empirical quantiles are shown along with 95\% confidence bands for each LIRADS score
    and controls. For cases, quantiles are shown across all lesional (red) and peri-lesional (blue) areas, and for controls, quantiles are shown across all liver masks (yellow). Distributions of log-lesion volumes are also shown for each LIRADS score.
    Observed lesion volumes ranged from 0.17-0.950 cm$^3$, with approximate diameters of 0.70-12.2 cm.}
\end{figure}

{\noindent \underline{Quantlet: Quantile-based Features based on EPM}:} Figure \ref{fig:Fig2} visualizes differences in the quantile distributions between different LIRADS scores as well as between cases vs controls. It is clear that the quantile distributions for the lesion pixels are often flatter compared to the distributions for non-lesional pixels. In particular, the non-lesional distribution has near-zero values and higher values corresponding to left and right tails respectively, which represents greater variability in the pixel distribution. These visual findings provide justification of incorporating quantile-based features, representing the distribution of pixel-level EPM values for diagnosis based on cancer imaging data.

We computed functional quantlet features that capture the information embedded in the distribution of EPM values across all pixels in the lesion and/or the liver (depending on the analysis goals). These functional features encompass a much richer set of information spanning the entire distribution compared to summary statistics typically reported in existing pipelines such as \cite{van2017computational}. The functional quantlet features rely on a basis expansion that enables smoothing and potential interpolation, thus reducing the impact of unequal-sized lesions. Further, they enable a parsimonious representation via a near-lossless property that ensures that only a moderate number of these quantlets are able to capture almost all of the information in the quantile distribution. 
The full mathematical details for deriving the quantlet features are described in the Supplementary Materials Section S.

{\noindent \underline{Classical radiomics features based on T1-MRI:}} In addition to EPM images, classical/structural radiomics features were also extracted based on the arterial phase of MRI scans. A standardized pipeline was used to extract these classical radiomics features that is described in the Supplementary Materials Section S2.

\subsection{Description of Analysis Goals} The first goal (Aim 1: Diagnostic Aim) involves developing an unsupervised classifier to diagnose cases vs controls using the EPM-derived enriched radiomics features at both pre-diagnostic and diagnostic visits. Control subjects had a single scan, while lesion cases had two, yielding 97 control and 73 lesional images. In Aim 1A, classification is performed using whole-liver EPM images without incorporating lesion masks or LIRADS scores when computing functional quantlet features. Aim 1B extends this by restricting the enriched radiomics features to lesional and peri-lesional regions. 
The second goal (Aim 2: Risk Stratification) focuses on lesion level classification, with the goal being to differentiate between aggressive tumors vs mild tumors by combining information across both visits. In order to simplify the analysis, we do not explicitly accounting for longitudinal dependence within subjects. A tumor is labeled as aggressive if it has LIRADS score $>3$, and mild otherwise. This results in a sample size of 84 for this analysis, with 39 tumors having LIRADS score higher than 3.  Aims 1b and 2 establish proof-of-concept of the proposed unsupervised diagnostic approach based on enriched radiomics features by comparing the performance against gold standard LIRADS scores.

Aim 3 extends beyond the cross-sectional analyses in Aims 1 and 2 to examine subject-specific longitudinal changes in liver EPM across visits, using the longitudinal Bayesian Tensor Response Regression (l-BTRR) model. The objective is to identify spatially localized EPM changes within and around lesion ROIs and assess whether these changes differ between aggressively progressing tumors (defined by a $\geq$ 2 increase in LIRADS) and stable or mildly aggressive tumors (no change or change of 1 in LIRADS score). This is done by comparing the proportion of lesion pixels showing significant positive or negative EPM changes between the two groups under the l-BTRR analysis. The analysis includes 26 lesions from 23 subjects (3 with two distinct baseline lesions and 20 with one), with 9 classified as aggressive (mean inter-visit duration: 325 days) and 17 as mild (mean: 312 days). 

\subsection{Statistical Model for Classification based on Enriched Radiomics Features}

We generalize the classical logistic regression models \cite{pan2021safe} to include functional predictors in the form of smoothed quantile distributions along with classical structural radiomics features. In particular, we a penalized logistic regression with functional predictors (corresponding to smoothed quantile curves) and additional scalar covariates (corresponding to classical structural radiomics and demographic features). Smoothed quantile distributions are computed based on the full image in Aim 1A, while they are computed separately for the lesional and the peri-lesional areas in Aim 1B, thus capturing rich distributional information across all pixels. 

We imposed $L_1$ penalty on the regression coefficients for the penalized logistic regression model that automatically down-weights the unimportant coefficients to zero, thus reducing overfitting and resulting in model parsimony. The regularization parameter $\lambda$ controlling the sparsity level was selected via cross-validation, and this choice of $\lambda$ is subsequently used in all model fitting. We denote the model as as scalar-on-functional quantile classification model that was implemented in R using the {\it glmnet} package. The full mathematical details of the model are provided in the Supplementary Materials.

\subsection{Statistical Model for Inferring Longitudinal Imaging Changes}

In addition to the prediction tasks based on enriched radiomics features, we also investigated whether the spatially distributed patterns of longitudinal changes in the EPM values could differentiate aggressively growing lesions from stable lesions. In our longitudinal analysis, aggressively growing lesions are denoted as those that show a LIRADS score change of two or more between visits, while stable tumors are those that show no LIRADS score change or a LIRADS score change of one across visits. In order to evaluate spatially distributed significant longitudinal EPM changes, we fit the longitudinal Bayesian tensor response regression (l-BTRR) model proposed in \cite{kundu2023bayesian} to the imaging data at baseline and follow-up visits. This model allows one to incorporate the spatial information in the image via a low-rank PARAFAC decomposition for the regression coefficient \cite{kundu2023bayesian}, results in spatial smoothing, and simultaneously results in massive dimension reduction in the number of model parameters.

The l-BTRR model was jointly fit across all imaging pixels, separately for each case subject. 
The BTRR model assumes a linear rate of change of EPM values across visits, and assumed spatially distributed imaging changes that are captured via tensor coefficients. The BTRR model is fit using an efficient Markov chain Monte Carlo (MCMC) sampler derived in \cite{kundu2023bayesian} that proceeds by sampling from full conditionals. Additional mathematical details about the modeling framework are provided in the Supplementary Materials.

\subsection{Benchmarking Comparisons}
One of the main goals of the study is to illustrate that the enriched radiomic characteristics produced a higher prediction accuracy compared to the classical radiomic characteristics used in the literature. Various competing models were considered that differed in terms of the sets of predictors used for classification. A leave-out-one cross-validation (LOOCV) strategy was used to evaluate the predictive performance under a penalized logistic regression approach. In particular, we compared with classification analyses involving: (i) demographic covariates only; (ii) summary measures such as mean or median of the pixel-level EPM values along with demographics; (iii) empirical quantiles along with demographics; (iv) summary structural radiomics features with demographics; and (v) the features considered in (ii)-(iii) along with structural radiomics features. The empirical quantiles are also considered probabilistic radiomic features but they differ from the proposed quantlet features that incorporate additional smoothing, resulting in more robust estimates. We consider various numbers of basis coefficients corresponding to classification analysis based on quantiles. 

Several metrics were used to evaluate the predictive performance. These measures included: (a) sensitivity that is defined as the power to detect true cases in Aim 1 and the power to detect tumors with higher grades in Aim 2; (b) the corresponding specificity rates that is equivalent to 1- false discovery rate; (c)  the F1 score that is the harmonic mean of sensitivity and specificity; ; (d) accuracy that is defined as the proportion of correctly identified cases and controls and (e) the area under the receiver operating characteristic curve (AUC) that is used to measure the overall quality of the classification. 

\section{Results}

The results for Aims 1 and 2 are presented in Tables \ref{tab:aim1} and \ref{tab:aim2}, and the results of Aim 3 are shown in Figures \ref{fig:propProg}. For the Aims 1 and 2, we report the results over varying number of basis coefficients included in the classification model, where these coefficients are included based on their ranking in a decreasing order of importance. 

\subsection{Diagnostic Classification}

From Table \ref{tab:aim1}, it is clear that the classification accuracy improves when using the enriched radiomics features along with demographics. In particular,
we observe that using the enriched features results in comparable or slightly improved sensitivity and specificity  compared to using structural radiomic features alone. The enriched radiomic analysis enhances the AUC compared to using either structural or probabilistic features alone, particularly in Aim 1B. The sensitivity and specificity under the enriched radiomics analysis were greater than 80\% and 90\% respectively, which is desirable, and indicates an overall improvement in diagnostic accuracy.

In contrast, using the mean and median summary statistics from the image results in poor performance, 
implying poor performance when using these summary measures that do not account for the heterogeneity in the image. Further, including only demographic features results in a zero AUC implying no classification ability. 
Table \ref{tab:aim1} shows that while the sensitivity is greater than 80\%  and specificity is higher than 90\% under both Aims 1A-1B, the performance is superior for Aim 1B in AUC. The improved performance for Aim 1B is due to the fact that the quantile distributions arising from the lesional and peri-lesional areas were considerably more different than the quantile distributions arising from healthy liver parenchyma. In contrast, the entire liver image was used for Aim 1A without any lesion level information that may have diluted the discriminative ability of the methods.


\begin{table}[h]
    \centering
    \begin{tabular}{|c|c|c|c|c|c|c|c|c|c|c|}
    \toprule
        & \multicolumn{10}{|c|}{Cases vs Control Diagnosis} \\
        & \multicolumn{5}{|c|}{Aim 1A} & \multicolumn{5}{|c|}{Aim 1B}\\
    \midrule
         Features & Sens & Spec & Acc & F1 & AUC & Sens & Spec & Acc & F1 & AUC \\ 
        \midrule
        D&0&1&0.5&NA&0&0&1&0.5&NA&0 \\
        Mean&0.21&0.91&0.56&0.32&0.57&0.21&0.91&0.56&0.32&0.57 \\
        Median&0.17&0.96&0.57&0.28&0.55&0.21&0.95&0.58&0.33&0.55 \\
        \midrule
        EQ(10)&0.24&0.95&0.6&0.37&0.56&0.24&0.95&0.6&0.37&0.56 \\
        EQ(30)&0.24&0.95&0.6&0.37&0.56&0.24&0.95&0.6&0.37&0.56 \\
        EQ(50)&0.24&0.95&0.6&0.37&0.55&0.24&0.95&0.6&0.37&0.56 \\
        \midrule
        SQ(10)&0&1&0.5&NA&0.50&0.86&0.77&0.81&0.79&0.85 \\
        SQ(30)&0.01&1&0.51&0.03&0.51&0.84&0.76&0.8&0.78&0.84 \\
        SQ(50)&0.01&1&0.51&0.03&0.51&0.86&0.73&0.79&0.77&0.83 \\
        \midrule
        R&0.83&0.93&0.88&0.86&0.92&0.83&0.93&0.88&0.86&0.92 \\
        \midrule
        Mean + R&0.8&0.96&0.88&0.86&0.92&0.8&0.96&0.88&0.86&0.92 \\
        Median + R&0.83&0.93&0.88&0.86&0.92&0.83&0.93&0.88&0.86&0.92 \\
        \midrule
        EQ(10) + R&0.79&0.95&0.87&0.85&0.92&0.81&0.95&0.88&0.86&0.92 \\
        EQ(30) + R&0.77&0.95&0.86&0.84&0.92&0.79&0.95&0.87&0.85&0.92 \\
        EQ(50) + R&0.79&0.95&0.87&0.85&0.92&0.76&0.95&0.85&0.83&0.91 \\
        \midrule
        SQ(10) + R&0.8&0.96&0.88&0.86&0.94&0.8&0.96&0.88&0.86&0.96 \\
        SQ(30) + R&0.8&0.96&0.88&0.86&0.94&0.8&0.96&0.88&0.86&0.96 \\
        SQ(50) + R&0.86&0.92&0.89&0.87&0.92&0.8&0.96&0.88&0.86&0.96 \\
        \bottomrule 
    \end{tabular}
    \caption{Aim 1 Results: Leave-out-one CV results for case/control classification. In Aim 1A, EPM features are taken from liver masks for control and cases, whereas for Aim 1B, EPM features for the cases are taken from their lesion masks. Different predictors were considered including Mean and Median EPM, EPM empirical quantiles (EQ), EPM quantlets (SQ), and PyRadiomics features (R). We considered varying number of quantiles and 70 PyRadiomics features (R).}
    \label{tab:aim1}
\end{table}

\begin{table}[h]
    \centering
    \begin{tabular}{|c|c|c|c|c|c|}
    \toprule
        & \multicolumn{5}{|c|}{Risk Stratification} \\
    \midrule
         Features & Sens & Spec & Acc & F1 & AUC \\ 
        \midrule
        D&1&0&0.5&0.7&0\\
        Mean&0.56&0.3&0.43&0.52&0.35\\
        Median&0.54&0.39&0.47&0.53&0.41\\
        \midrule
        EQ(10)&0.56&0.49&0.52&0.56&0.50\\
        EQ(30)&0.46&0.46&0.46&0.48&0.42\\
        EQ(50)&0.54&0.42&0.48&0.53&0.43\\
        \midrule
        SQ(10)&0.67&0.49&0.58&0.63&0.58\\
        SQ(30)&0.56&0.42&0.49&0.55&0.50\\
        SQ(50)&0.69&0.3&0.5&0.61&0.50\\
        \midrule
        R&0.8&0.64&0.72&0.76&0.79\\
        \midrule
        Mean + R&0.8&0.64&0.72&0.76&0.79\\
        Median + R&0.8&0.64&0.72&0.76&0.79\\
        \midrule
        EQ(10) + R&0.87&0.64&0.75&0.8&0.81\\
        EQ(30) + R&0.87&0.64&0.75&0.8&0.81\\
        EQ(50) + R&0.87&0.64&0.75&0.8&0.81\\
        \midrule
        SQ(10) + R&0.85&0.73&0.79&0.82&0.87\\
        SQ(30) + R&0.85&0.73&0.79&0.82&0.85\\
        SQ(50) + R&0.85&0.73&0.79&0.82&0.86\\
        \bottomrule
    \end{tabular}
    \caption{Aim 2 Results: Leave-out-one CV results for cross-sectional classification (Aim 2) of tumor grades of aggressive (LIRADS score greater than three) versus mild (LIRADS score of three or below).
    }  
    \label{tab:aim2}
\end{table}

\subsection{Lesion Score Stratification}
The advantages of the enriched radiomic features become more evident when classifying the LIRADS scores
in Aim 2 (see Table \ref{tab:aim2}). In particular, the AUC under the enriched radiomic features is considerably higher than competing analytical methods. In particular, it registers consistent improvements across all performance metrics considered, highlighting the benefits of including the probabilistic smoothed quantlet features over the less robust empirical quantile features. Moreover, the classical structural radiomics features perform poorly in Aim 2, producing lower sensitivity and considerably poor specificity, and ultimately resulting in an AUC less than 0.8.  These results point to the importance of combining functional quantile-based and structural radiomic features  to achieve desirable lesion score stratification that is needed to identify advanced tumors.  Notably, we anticipate the classification AUC to increase to greater than 95\% for larger sample sizes, given that an AUC of 87\% was achieved for a moderate sample size used for our analysis.

\subsection{Longitudinal Changes in EPM Image is Correlated with Tumor Progression}

\begin{figure}
    \centering
    \includegraphics[width=0.8\linewidth]{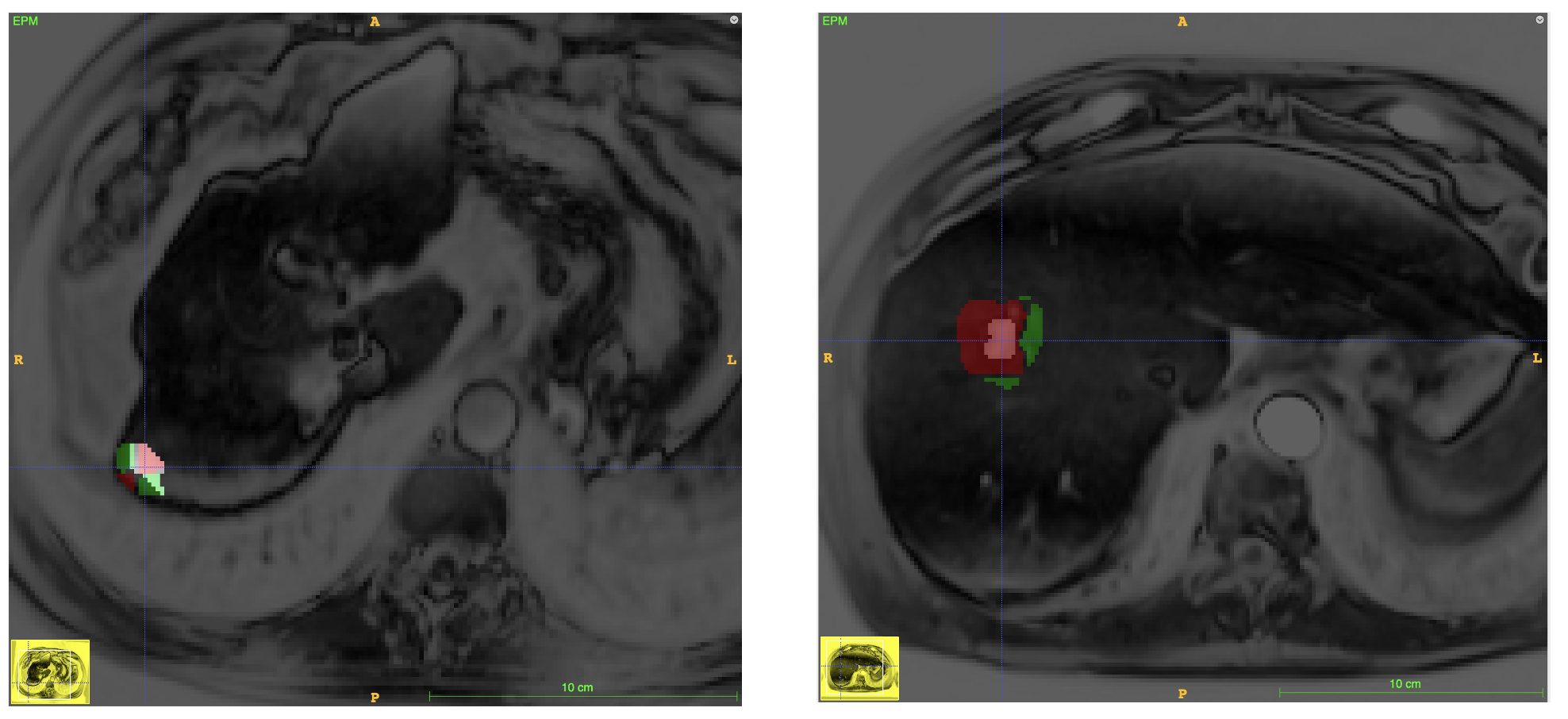}
    \caption{Example of subject-specific liver map showing regions of significant longitudinal EPM changes around lesion, as determined via longitudinal BTRR. White indicates presence of lesion, red indicates positive (+) EPM change, and green indicates negative (-) EPM change.}
    \label{fig:longEPM}
\end{figure}

\begin{figure}[h]
    \centering
    \includegraphics[width=1\linewidth]{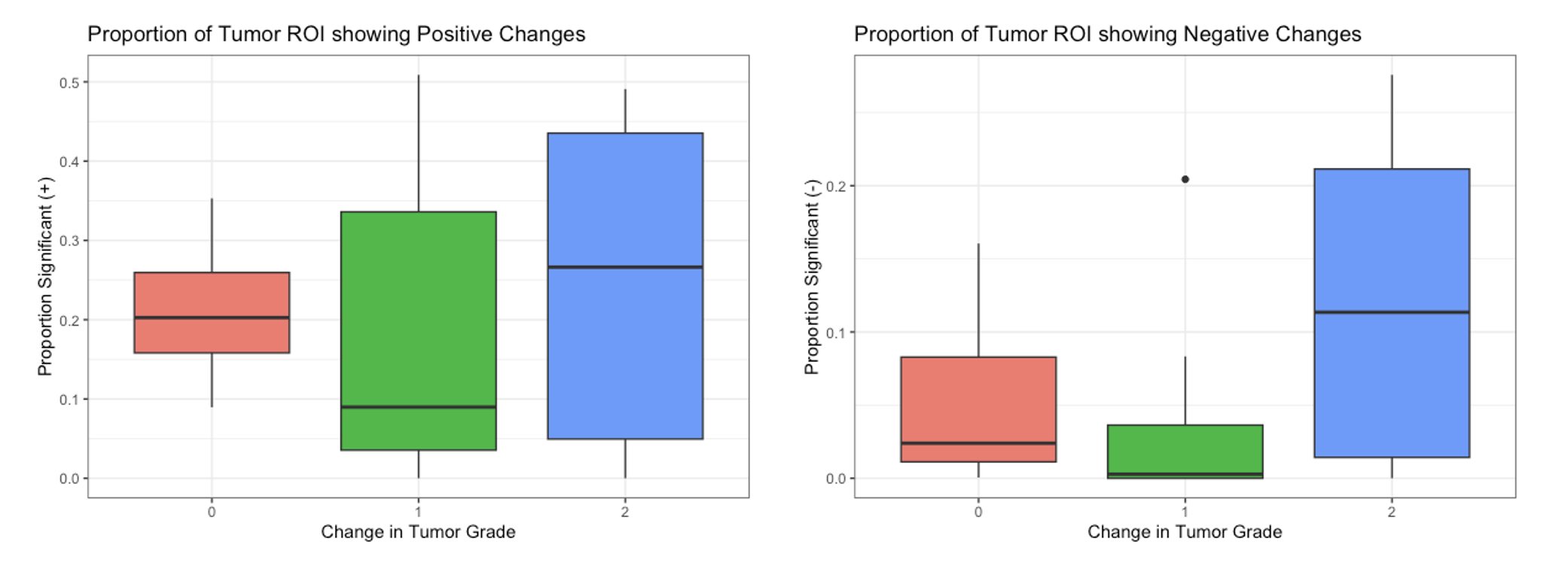}
    \caption{Boxplots displaying proportions of significant positive (left) and negative (right) EPM changes within lesion ROI, stratified by change in lesion score. For right-hand plot, the medians are $0.0240$ for no change, $0.00278$ for change of $+1$, and $0.113$ for change of $+2$; the proportions of lesions with $<5\%$ negative significant changes are $0.667$ for no change, $0.727$ for $+1$ change, and $0.333$ for $+2$ change.
    }
    \label{fig:propProg}
\end{figure}

Our l-BTRR analysis revealed spatially localized, significant longitudinal changes in EPM values within and around lesion ROIs. Figure \ref{fig:longEPM} illustrates examples of both positive and negative EPM changes over time for two individuals with LIRADS score of three. Boxplots in Figure \ref{fig:propProg} show how the proportion of significantly changing voxels within each ROI varies by change in LIRADS score across visits. Overall, we observed that positive EPM changes were more common than negative ones. For aggressive tumors, the median proportion of voxels showing significant EPM increases exceeded 0.25, compared to $\leq$0.2 for stable tumors. However, aggressive lesions also showed a higher median proportion of voxels with significant EPM reductions (0.12) than stable or mildly aggressive lesions.

These findings suggest that a high proportion of voxels with significant EPM reductions may indicate future tumor progression, whereas smaller reductions may be associated with more stable growth. While aggressive tumors tend to show greater longitudinal EPM change overall compared to stable tumors, the distinction between aggressive vs stable tumors is most visible with respect to significant spatial decreases in EPM values over longitudinal visits. This pattern may offer clinicians a noninvasive and clinically useful imaging-based prognostic biomarker to support early identification of aggressive lesions.

\section{Discussion}

Our results demonstrate that EPM-derived features improve performance in diagnostic, lesion risk stratification, and longitudinal liver imaging tasks. For diagnostic and lesion-level tasks, models incorporating quantlet-based representations of the EPM distributions achieved better performance than those relying on structural radiomics alone or empirical EPM quantiles. This indicates that stabilizing and smoothing the EPM distribution enables more reliable extraction of enhancement-related information that complements traditional MRI-derived features. In the longitudinal setting, the tensor-based analysis revealed clear differences in voxel-level EPM changes between more progressive and less progressive lesions, suggesting that temporal patterns in enhancement deviations provide additional information beyond what is available from static imaging. Together, these findings highlight the value of incorporating EPM-driven features for improved lesion characterization across both cross-sectional and longitudinal imaging contexts.


The improved performance of the cross-sectional tasks indicates that EPM-derived features provide information that complements the structural radiomics extracted directly from the MRI. Although structural radiomics captures shape and textural attributes, many handcrafted features are known to lack robustness for tumor classification tasks due to sensitivity to segmentation and acquisition variability \citep{Robinson2019-ta}. In addition, inherent noise and variability in MRI-based cancer imaging reduce the signal-to-noise ratio, further limiting the reliability of classical radiomics features for assessing lesion severity \citep{Cattell2019-ro}. The empirical EPM quantiles also showed limited performance, likely due to voxel-level fluctuations and the effects of small lesion size on distributional stability \citep{yang2019quantile}. In contrast, the quantlet representation provided a smoothed and low-dimensional summary of the EPM distribution that reduced noise-driven variability and enabled the classifier to capture enhancement heterogeneity more reliably. This explains why quantlet-based features consistently outperformed both empirical quantiles and structural radiomics, and why combining structural and quantlet features yielded the strongest overall performance: each contributes distinct and complementary information about lesion appearance.

The longitudinal analysis from Aim 3 further demonstrates that temporal patterns in the EPM signal provide meaningful information about lesion evolution that is not captured by cross-sectional features alone. The tensor-based model identified clear differences in voxel-level EPM changes between lesions, with more progressive lesions showing a greater proportion of significant decreases in EPM values over time. These results suggest that longitudinal shifts in enhancement deviations reflect evolving imaging characteristics that may be relevant during follow-up.
By aggregating spatial information and stabilizing comparisons across visits, the tensor approach enabled consistent detection of these temporal patterns despite variability in acquisition or lesion size. This highlights the value of incorporating longitudinal imaging data into radiomics workflows, especially in surveillance settings where early changes are often difficult to quantify using static measurements alone.

Several considerations should be noted. Smaller lesions inherently provide fewer informative voxels, which may limit the stability of distribution-based features even with quantlet smoothing. Segmentation accuracy and the delineation of peri-lesional regions may influence results and should be standardized in future studies. Because this work was conducted at a single institution, external validation across scanners, protocols, and patient populations is necessary to establish generalizability. Integrating EPM-based features with additional imaging biomarkers or clinical covariates may further enhance classification performance.

Taken together, these results highlight the potential of stabilizing EPM-based features to strengthen radiomics pipelines and improve the characterization of lesion appearance over time. By combining a functional representation of EPM with a spatially structured longitudinal model, the proposed framework offers a flexible foundation for analyzing heterogeneous imaging patterns. The underlying principles of smoothing distributional voxel-level features and modeling temporal changes in a high-dimensional spatial field are broadly applicable and can be adapted to radiomics studies in other cancer imaging domains. Future investigations involving additional tumor types and imaging modalities will be useful for assessing generalizability and for integrating these approaches into multi-modality imaging workflows.






\vspace{6pt} 

\supplementary{The following supporting information can be downloaded at:  \linksupplementary{s1}, Figure S1: title; Table S1: title; Video S1: title.}




\authorcontributions{
conceived the analysis, S.K., E.K., and A.R. ; helped collect the data, E.K., P.J.; helped collect and pre-process the data, R.H., G.J., M.R., M,B., P.P, L,B., D.F.; performed the numerical analyses, A.R. and T.Y.; write and edit the manuscript, A.R., T.Y., and S.K. helped, helped edit the manuscript, E.K. All authors have read and agreed to the published version of the manuscript.}

\funding{This research received no external funding.
This research was funded by grants R01AG071174 and R01MH120299. The findings represented in this article do not represent official NIH views.
}

\institutionalreview{
The study was conducted in accordance with the Declaration of Helsinki, and approved by the Institutional Review Board of The University of Texas MD Anderson Cancer Center (PA14-0646) with latest approval on 4 October 2022.
}

\informedconsent{
Patient consent was waived by the IRB due to the study being retrospective.

}

\dataavailability{
The data was obtained from Baylor College of Medicine. The datasets generated and/or analyzed during the current study are not publicly available due to ongoing research analysis but are available from the corresponding author on reasonable request.}

\conflictsofinterest{
The authors declare no conflicts of interest.
} 

\begin{adjustwidth}{-\extralength}{0cm}

\reftitle{References}


\bibliography{literature}

\PublishersNote{}
\end{adjustwidth}
\end{document}


\flushbottom
\maketitle
%
%

\section{Summary of lesion level information}
The following Table describes lesion level information. 
\begin{table}[h]
    \centering
    \small
    \begin{tabular}{lccccc}
    \toprule
        &\textbf{ LIRADS Score = 2} &\textbf{ LIRADS Score = 3} &\textbf{ LIRADS Score = 4} &\textbf{ LIRADS Score = 5} &\textbf{ Overall}\\
        &\textbf{ (N=6)} &\textbf{ (N=42)} &\textbf{ (N=17)} &\textbf{ (N=23)} &\textbf{ (N=88)}\\
    \midrule
    \textbf{Mean EPM in Lesion} & &  &  &  & \\
        \quad Mean (SD) & 1.05 (0.370) & 0.654 (0.417) & 0.539 (0.341) & 0.678 (0.347) & 0.665 (0.394)\\
        \quad Median [Min, Max] & 1.19 [0.379, 1.39] & 0.528 [0.187, 1.78] & 0.425 [0.170, 1.48] & 0.604 [0.219, 1.61] & 0.557 [0.170, 1.78]\\
        \addlinespace
        \textbf{Mean EPM around Lesion}  & &  &  &  & \\
        \quad Mean (SD) & 0.543 (0.309)& 0.557 (0.363) & 0.446 (0.290) & 0.643 (0.347) & 0.557 (0.343)\\
        \quad Median [Min, Max] & 0.562 [0.134, 0.902] & 0.437 [0.137, 1.37] & 0.334 [0.170, 1.22] & 0.544 [0.214, 1.44] & 0.474 [0.134, 1.44]\\
        \addlinespace
        \textbf{Number Of Voxels in Lesion} & &  &  &  & \\
        \quad Mean (SD) & 6500 (12700) & 346 (472) & 735 (622) & 26600 (113000) & 7710 (58000)\\
        \quad Median [Min, Max] & 321 [176, 32100] & 220 [32, 3090] & 575 [200, 2380] & 972 [69, 544000] & 356 [32, 544000] \\
        \addlinespace
        \textbf{Lesion Volume} & & & & & \\
        \quad Mean (SD) & 12.4 (18.5)& 1.23 (0.961) & 2.78 (2.62) & 53.2 (198) & 15.9 (102)\\
        \quad Median [Min, Max]  & 0.546 [0.299, 36.9] & 0.937 [0.178, 4.57] & 1.58 [0.335, 9.40] & 4.53 [0.422, 954] & 1.47 [0.178, 954] \\
        \addlinespace
        \textbf{Lesion Diameter} & & & & & \\
        \quad Mean (SD) & 1.99 (1.64) & 1.26 (0.311) & 1.61 (0.475) & 2.76 (2.34) & 1.77 (1.42) \\
        \quad Median [Min, Max] & 1.01 [0.830, 4.13] & 1.21 [0.698, 2.06] & 1.45 [0.862, 2.62] & 2.05 [0.931, 12.2] & 1.41 [0.698, 12.2]\\
    \bottomrule 
    \end{tabular}
    \caption{Summary of liver image information of lesions.}
    \label{tab:Summary_Lesion}
\end{table}

\section{Details on Extraction on Classical Radiomics Features}

For controls, each ROI (6mm in diameter) was manually defined over background liver parenchyma, and subsequently the structural radiomics features were extracted from these ROIs using the {\it Pyradiomics} package \cite{van2017computational}. For cases, the structural radiomics features were computed separately for each lesion as well as for an independent non-lesional (pre-defined) ROI. Both liver masks and lesion masks were used for defining ROIs, where the lesion segmentation was conducted based on the arterial phase of the T1w-MRI scan. When computing the structural radiomics features for cases, we did not differentiate between multiple lesions with the same LIRADS score.

From a total of 108 radiomics features extracted from suitable ROIs, we included four categories of structural radiomics information: 24 Gray Level Co-occurrence Matrix (GLCM) features, 16 Gray Level Run Length Matrix
(GLRLM) features, 16  Gray Level Size Zone Matrix (GLSZM) features, and 14 shape-based features. We excluded variables representing first-order structural radiomic information (e.g. energy, entropy, interquartile range) due to potential overlap with functional radiomic information.

\section{Details about Derivation of Functional Radiomics Features}

\subsection{Methodology for Extracting Probabilistic Radiomics Features via Quantlet Basis Approach}
The starting point for the proposed scalar-on-functional quantile classification approach is to compute the empirical quantiles. The quantile distributions naturally account for heterogeneity in the image (arising from lesional, peri-lesional, and healthy regions in the organ) and is able to tackle different image sizes across samples. Figure 2 in the Main Paper visualizes differences in the quantile distributions between different LIRADS scores as well as between cases vs controls. It is clear that the quantile distributions for the lesion pixels are often flatter compared to the distributions for non-lesional pixels. In particular, the non-lesional distribution has near-zero values and higher values corresponding to left and right tails respectively, which represents greater variability in the pixel distribution.

First, we demonstrate how to derive the empirical quantiles from the EPM images. Denote the $i$th image as $X_i$ comprising $m_i$ pixels ($i=1,\ldots,N$). We note that $X_i$ may include the full liver image for subject level classification and a smaller image involving lesional and per-lesional areas for lesion level classification. Further, denote the $m_i$ empirical quantiles for the $i$th sample as $\{Q_i(p_{1}),\ldots,Q_i(p_{m_i})\}$, which are derived from $X_i$, $i=1,\ldots,N,$ for $0<p_j<1$. Mathematically, the quantile function is defined as $Q_X(p) = F^{-1}_X(p) = inf\{x: F_X(x)\ge p \}, \mbox{ } 0\le p \le 1, $
where $inf(\cdot)$ denotes the infimum, and $F_X$ denotes the cumulative distribution function, i.e. $p=F_X(x) = P(X\le x)\in [0,1]$. Further, denote the order statistics from the $i$th image as $\{X^{(1)}_i, \ldots, X^{(m_i)}_i \}$ that are arranged in an increasing order. We note that for $p\in [1/(m_i+1), m_i/(m_i+1) ]$, the empirical quantile function for the $i$th sample is computed as 
\begin{eqnarray}
\hat{Q}_i(p) = (1-w_i)X_i^{((m_i+1)p)} + w_i X_i^{((m_i+1)p+1)}, \label{eq:emp_quantile}
\end{eqnarray}
where $w_i = X_i^{((m_i+1)p+1)}- X_i^{(m_i+1)p}$. The above expression is obtained via interpolation, where $[x]$ is an integer $\le x$. In practice, we will focus our modeling on quantiles lying within $[\delta,1-\delta]$ where $\delta = \max_{i\le n} \{ 1/(m_i+1) \}$.

We note that the empirical quantile function can be computed on a grid $\{p_1,\ldots,p_K\}$ lying within $[\delta,1-\delta]$ using (\ref{eq:emp_quantile}), where the grid is standardized across images. Subsequently, the corresponding empirical quantiles $\{Q_X(p_1),\ldots,Q_X(p_K) \}$ could be potentially used as covariate features for prediction. However, the quality of the empirical quantiles computed over the standarized grid will depend on the size of the images. While it may be possible to estimate these quantiles with reasonable accuracy corresponding to larger images with several thousands of pixels, the same may not be true for smaller images with fewer pixels or small lesions. This is due to the fact that the order statistics $\{X^{(1)}_i, \ldots, X^{(m_i)}_i \}$ are potentially spaced farther apart for images with fewer pixels, that results in an inferior interpolation in (\ref{eq:emp_quantile}). Therefore, directly using the empirical quantile distribution as functional predictors may not be desirable, given that their accuracy depends on the image size and other factors such as lack of smoothness across neighboring quantiles.

A more appealing approach is to propose a statistical smoothing method over the empirical quantiles with the goal being to provide more accurate and robust quantile-based features. As shown in \cite{yang2020quantile}, it is possible to find an empirical set of basis functions $\{ \psi_1, \ldots, \psi_K\}$ on the quantile space to approximate the empirical quantiles with extremely high accuracy (a property known as near lossless). That is,  
\begin{eqnarray}
Q_{X_i}(p_j) \approx \sum_{k=1}^K Q^*_{ik} \psi_k(p_j), \label{eq:basis_expansion}
\end{eqnarray}
where $Q^*_{ik}$ represent the basis coefficients for the $k$th basis function for the $i$th sample, and $\{ \psi_1(p),\ldots, \psi_K(p)\}$ represent a set of orthogonal quantlet basis functions. Under suitable choice of basis functions, even a small to moderate number of basis functions in (\ref{eq:basis_expansion}) is sufficient to guarantee an accurate approximation, as shown in \cite{yang2020quantile}. 
In contrast to the empirical quantiles that is derived directly from the imaging data, the functional quantlet approach uses a wavelet-based smoothing approach that leads to information sharing across neighboring quantile levels. Therefore, empirical quantiles are more susceptible to noise in the image and the imaging size compared to the functional quantlet-based approach.

The basis coefficients in (\ref{eq:basis_expansion}) are obtained in a straightforward manner as follows. Let us denote $\boldsymbol {Q_i} = (Q_i(1),\ldots,Q_i(m_i))$, and $\boldsymbol\Psi $ be a $K\times m_i$ with the $\boldsymbol\Psi(k,j) = \psi_k(p_j)$. Then the basis coefficients are given as $\boldsymbol  Q^*_i = (Q^*_{i1}, \ldots, Q^*_{iK}) = \boldsymbol Q_i \boldsymbol\Psi^T_i (\boldsymbol\Psi_i \boldsymbol\Psi^T_i)^{-1}$, where $K<< \min\{m_1,\ldots, m_N\}$ in our settings involving a large number of pixels in the image. We note $\boldsymbol Q^*_i$ contains virtually all of the information in the raw data $\boldsymbol Q_i$ via the near lossless property, but it is simultaneously more robust to noise and overfitting in the empirical quantile distribution. Therefore, we use the smoothed quantile function $\hat{Q}_{X_i}(p_j) = \sum_{k=1}^K \hat{Q}^*_{ik} \psi_k(p_j)$ as the functional predictor in our model in lieu of the empirical quantile distribution, where $\hat{Q}^*$ denotes the estimated coefficients under (\ref{eq:basis_expansion}). More details on such coefficient estimation as well as the choice of $K$ are provided in the following subsection.

\subsection{Choice of Quantlet Basis}
We adapt the approach in \cite{yang2020quantile} for choosing the basis functions, and design quantlets as a suitable set of basis functions that were shown to work well for representing the quantile distribution. We create empirical quantlet bases from a given dataset, serving as a robust representation of each subject's quantile function. The basis function is able to reconstruct the quantile distribution with greater than $99.9\%$ accuracy using a moderate number coefficient terms. This type of low dimensional feature extraction results in massive dimension reduction for large images and yields a more faithful reconstruction of the quantile distribution compared to alternate representations such as Legendre-polynomials \cite{ghosal2023distributional} that can result in poor tail approximation for heavy tailed distributions \cite{hosking1990moments}. These bases are employed in our subsequent quantile functional regression model. 

Consider the set of quantile functions $\mathcal{Q}$ such that $\mathcal{Q}:= \{ Q:\int \big(Q^2(p) d\Pi(p) \big)^{1/2} < \infty \}$, where $\Pi(\cdot)$ is a uniform density with respect to the Lebesgue measure. Define the first two basis functions to be a constant basis $\zeta_1(p)=1$ for $p\in [0,1]$ and standard normal quantile function $\zeta_2(p)=\Phi^{-1}(p)$. These orthogonal bases cover the entirety of Gaussian quantile functions, where the initial coefficient signifies the mean and the second coefficient represents the distribution's variance. An overcomplete dictionary is created, comprising these bases as well as numerous elements generated from Beta cumulative density functions (CDFs). The additional basis functions are constructed as $\zeta_j(p) = P_{N^{\perp}}(\frac{F_{\theta_k}(p) - \mu_{\theta_k}}{\sigma_{\theta_k}}), j\ge 3,$
where $F_{\theta_k}$ is the cumulative distribution function (CDF) of a Beta$(\theta_k)$ distribution for some positive parameters $\theta_k = \{a_k, b_k \}$, $\mu_{\theta_k}=\int_{0}^1 F_{\theta_k}(u) du$, and $\sigma^2_{\theta_k} = \int_{0}^1 (F_{\theta_k}(u) - \mu_{\theta_k})^2 du$ are the centered and scaled values of these distributions for standardization, respectively, and $P_{N^{\perp}}$ indicates the projection operator onto the orthogonal complement to the Gaussian basis elements $\zeta_1$ and $\zeta_2$. In other words, $P_{N^{\perp}}(f(p)) = f(p) - \zeta_1(p) \int_{0}^1 f(u)\zeta_1(u) du - \zeta_2(p) \int_{0}^1 f(u)\zeta_2(u) du $.

Put together, the set $\mathcal{D}^O = \{ \zeta_1 \cup \zeta_2\} \cup \{ \zeta_k: \theta_k \in \mathcal{\theta} \}_{k=1}^{K_0}$ represents an overcomplete dictionary, where  $\mathcal{\theta} $ represents the parameter space of interest. In practice, to fix the number of dictionary elements, we choose a grid on the parameter space to obtain $\mathcal{\theta} = \{\theta_k = (a_k,b_k) \}_{k=3}^{K_0}$ by uniformly sampling on $\mathcal{\theta}\subset (0,J)^2$ for some sufficiently large $J$, and and choosing $K_0$ to be a large integer. More details on how to choose the parameter space is provided in \cite{yang2020quantile}, along with additional theoretical results that justify the choice of a large dictionary of Beta CDF to represent any quantile function whose first derivative is absolutely continuous.

Once this overcomplete basis set has been constructed, a sparse subset of basis functions are chosen that can represent the quantile functions across all samples. This is made possible by first fitting a Lasso regression \cite{tibshirani1996regression} to select a sparse set of wavelets for each subject separately, and subsequently taking the union across these sparse sets of basis elements across all subjects, to obtain a unified set of dictionary elements $\mathcal{D}$. Subsequently, a leave-one-out concordance correlation coefficient (LOOCCC) is proposed that allows one to reduce the number of basis elements in $\mathcal{D}$ even further, with minimal loss of information. This property is known as near lossless property. This goodness of fit criteria for the $i$th subject is given as $\rho_{(i)} = \frac{Cov(Q_i(\cdot),\sum_{k\in \mathcal{D}_{(i)}^U} \zeta_k(\cdot) Q^U_{ik})}{Var(Q_i(\cdot)) + Var(\sum_{k\in \mathcal{D}_{(i)}^U} \zeta_k(\cdot) Q^U_{ik})} $, where $\mathcal{D}_{(i)}^U$ is the set of the sparse basis elements obtained under the Lasso regression, combined across all but the $i$-th subject. Subsequently one can use $\rho^0= \min \{ \rho_{(i)}\}_i$ as an overall goodness of fit measure that can be used to select a small to moderate number of basis elements for our downstream prediction model, with no more than minimal loss of information  For example, in the Liver EPM dataset, we find that $13$ basis elements are able to preserve a concordance of at least 0.999, when reconstructing the quantile distributions from the entire liver mask. 
Once a reduced set of basis elements are obtained, subsequent Gram-Schimdt orthogonalization steps are used followed by a wavelet-based denoising step \cite{donoho1995wavelet}.

\section{Details of Scalar-on-functional quantile Classification Model}

 Let $y_i\sim Ber(p_i(Q_x))$ denote the binary outcome of interest, which follows a Bernoulli distribution with probability $p_i(Q_x)$ that depends on the quantile distribution of the image. Further, let ${\bf y} = (y_1,\ldots, y_N)$ denote the vector of binary outcomes, ${\bf w}$ denotes the demographic features, and ${\bf z}$ denotes the structural radiomics features derived from the arterial phase of the T1w-MRI scans. Let $\mathcal{X}
_i$ correspond to the full liver image for Aim 1A, and let it correspond to pre-defined ROIs encompassing the lesion for the cases in Aim 1B and for the lesion level analyses in Aim 2. When lesion segmentation is available, denote the lesion mask for the $i$th image as $\mathcal{L}_i$, so that $\mathcal{L}_i^c = X_i \setminus \mathcal{L}_i$ denotes the peri-lesional areas adjacent to the lesion. The likelihood is $L({\bf y}\mid \mathcal{X}) = \prod_{i=1}^N p^{y_i}_i(Q_x)(1-p_i(Q_x))^{1-y_i}$, where
\begin{align}
    & \frac{p_i(Q_x)}{1-p_i(Q_x)}  = \alpha + ({\bf w}_i^T, {\bf z}_i^T)\bm{\zeta} + \int_{0}^1 \hat{Q}_{\mathcal{X}_i}(p) \beta_{\mathcal{X}}(p)dp \nonumber \\
    & \approx \alpha + ({\bf w}_i^T, {\bf z}_i^T)\bm{\zeta}+ \sum_{k=1}^K \hat{Q}^*_{\mathcal{L}_ik}\gamma_k + \sum_{k=1}^K \hat{Q}^*_{\mathcal{L}_i^c,k}\eta_k , \label{eq:like}
\end{align}
\noindent where $\int_{0}^1 \hat{Q}_{\mathcal{X}_i}(p) \beta_{\mathcal{X}}(p)dp = \int \hat{Q}_{\mathcal{L}_i}(p) \beta_{\mathcal{L}}(p)dp + \int \hat{Q}_{\mathcal{L}^c_i}(p) \beta_{\mathcal{L}^c}(p)dp \approx \sum_{k=1}^K \hat{Q}^*_{\mathcal{L}_ik}\gamma_k + \sum_{k=1}^K \hat{Q}^*_{\mathcal{L}_i^c,k}\eta_k$,  $\beta(p) $ denotes the functional regression coefficient that can be represented as the basis expansion $\beta(p)\approx \sum_{k=1}^K \eta_k \psi_k(p) $, $Q^*$ are defined as the functional quantile features,${\bm \zeta}$ denotes the effects corresponding to the demographic features, and $\bm{\gamma}$ denotes the effects for the structural radiomics features.  Model (\ref{eq:like}) accounts for heterogeneity by assigning separate effects for the lesional and peri-lesional areas, and via smoothed quantile curves capturing rich distributional information across all pixels. We fit the following penalized likelihood criteria to estimate the model parameters in (\ref{eq:like}):
\begin{eqnarray}
    \hat{\bm{\theta}} = \mbox{arg min}_{\bm{\theta}} - \log(L({\bf y}\mid \mathcal{X},\bm{\theta}) + \lambda  |\bm{\theta}|_1, \label{eq:penlike}
\end{eqnarray}
where $\bm{\theta}=(\alpha, \bm{\zeta,\gamma,\eta})$ denotes the collection of model parameters, $\lambda$ denotes the penalty parameter for the $L_1$ penalty, with a larger value indicating greater sparsity and vice-versa. The imposed $L_1$ penalty down-weights the unimportant coefficients to zero, thus reducing overfitting and resulting in model parsimony. The penalized logistic regression model in  (\ref{eq:like})-(\ref{eq:penlike}) was implemented in R using the {\it glmnet} package.  The regularization parameter $\lambda$ was selected via cross-validation, and this choice of $\lambda$ is subsequently used in all model fitting. We denote the proposed model in (\ref{eq:like})-(\ref{eq:penlike}) as scalar-on-functional quantile classification model that is used to  address the following analysis goals.

\section{Details for the Bayesian Tensor Response Regression Model}

In order to focus on the longitudinal changes in EPM around the lesion areas that is of primary interest, we first use k-means segmentation to identify the spatially distinct lesioned areas. Subsequently, we limited our investigation to EPM changes within spherical regions of interest (ROI) centered around the baseline lesions with approximately twice the volume of the lesion. 

We write the l-BTRR model as follows for the $t$-th timepoint, and $l$-th subject-specific lesion for a given subject:
\begin{align}
    & \mathcal{Y}_{tl} = \mathcal{M}_l + \Gamma_l \mathcal{T}_{t} + \mathcal{E}_{tl}, \quad \epsilon_{tl}(v) \sim N(0, \sigma_{l}^2),  \quad  
    \mathcal{M}_{l} = \sum_{r=1}^R \mu_{1\bullet,r} \circ \mu_{2\bullet,rl} \circ \mu_{3\bullet,rl}, \quad 
     \Gamma = \sum_{r=1}^R \gamma_{1\bullet,rl} \circ \gamma_{2\bullet,rl} \circ \gamma_{3\bullet,rl}, \label{eq:mod1}
\end{align}
\noindent where $\mathcal{Y}_{tl} \in \mathbb{R}^{p_1 \times p_2 \times p_3}$ is the observed EPM image within the $l$-th lesion ROI mask for subject $i$ and timepoint $t$, $\mathcal{T}_{t}$ is the time after baseline for the $i$-th subject and $t$-th timepoint, and $\mathcal{M}_{l}$ and $\Gamma_{l}$ are tensor-valued coefficients (with rank $R$) representing the baseline intercept and time-slope, respectively. The error term, $\mathcal{E}_{tl}$ has the same dimensions as $\mathcal{Y}_{tl}$, and its $v$-th spatial element, $\mathcal{E}_{tl}(v) = \epsilon_{tl}(v)$, is assumed to follow a normal distribution with zero mean and variance $\sigma_{l}^2$. Although the tensor rank $R$ may be assumed to be distinct for the intercept and slope parameters, we choose a common value of $R$ for simplicity. The choice of rank $R$ is usually determined using a goodness of fit criteria such as DIC \cite{guhaniyogi2021bayesian}.

The above model assumes a linear rate of change in EPM values between visits and assumed spatially distributed imaging changes that are captured through tensor coefficients $\Gamma$. On the other hand, the intercept term $\mathcal{M}$ captures the spatial dependence in the image under a low-rank PARAFAC decomposition. Specifically, we impose a rank-$R$ PARAFAC decomposition on the tensor coefficients that expresses the coefficients as an outer product of low tank tensor margins, resulting in spatial smoothing and a massive dimension reduction from $p_1 \times p_2 \times p_3$ to $R(p_1+p_2+p_3)$ parameters. These factors results in robust estimates of longitudinal changes even when fitting the model separately for each case subject under the l-BTRR approach. We fit Model \ref{eq:mod1} using an efficient Markov chain Monte Carlo (MCMC) sampler derived in \cite{kundu2023bayesian} that proceeds by sampling from full conditionals. Subsequently, we perform inference on the slope parameters $\Gamma_{il}$ using joint credible intervals \cite{crainiceanu2007spatially} to discover imaging pixels with significant longitudinal EPM changes.



\bibliography{literature}